\documentclass[showpacs,aps,graphicx]{revtex4}%and
\usepackage{graphicx}

\begin{document}
\title{The heralded amplification for the single-photon entanglement of the time-bin qubit}

\author{Lan Zhou$^{1,2}$, Yu-Bo Sheng,$^{2,3}$\footnote{Email address:
shengyb@njupt.edu.cn}  }
\address{
 $^1$ College of Mathematics \& Physics, Nanjing University of Posts and Telecommunications, Nanjing,
210003, China\\
 $^2$Key Lab of Broadband Wireless Communication and Sensor Network
 Technology, Nanjing University of Posts and Telecommunications, Ministry of
 Education, Nanjing, 210003, China\\
$^3$Institute of Signal Processing  Transmission, Nanjing
University of Posts and Telecommunications, Nanjing, 210003,  China\\}

\begin{abstract}
We put forward an effective amplification protocol for protecting the single-photon entangled state of the time-bin qubit. The protocol only requires one pair of the single-photon entangled state and some auxiliary single photons. With the help of the 50:50 beam splitters, variable beam splitters with the transmission of $t$ and the polarizing beam splitters, we can increase the fidelity of the single-photon entangled state under $t<\frac{1}{2}$. Moreover, the encoded time-bin information can be perfectly contained. Our protocol is quite simple and economical. More importantly, it can be realized under current experimental condition. Based on the above features, our protocol may be useful in current and future quantum information processing.
\end{abstract}
 \maketitle

\section{Introduction}
Entanglement of flying qubits is a fundamental principle of
quantum information. It has been widely used in quantum communication
and quantum computation protocols over the past
decades, such as quantum teleportation \cite{teleportation}, quantum key distribution (QKD) \cite{qkd},  quantum secret sharing (QSS) \cite{qss}, quantum secure direct communication (QSDC) \cite{qsdc1,qsdc2}, quantum repeaters \cite{repeater1,repeater2}, and one-way quantum computation \cite{oneway,zhoucluster,shengcluster}.

Among various entanglement forms, the time-bin entanglement, which is a coherent superposition of single photon states in two or more different temporal modes is a simple and conventional form. For example, suppose a single-photon pulse enter two channels, where channel length difference (long-short) is much longer than the pulse duration. The output state consists of two well separated pulses with different times of arrival, and the quantum information is encoded in the arrival
time of the photon. We denote the two states as $|Long\rangle$ ($|L\rangle$) and $|Short\rangle$ ($|S\rangle$), respectively, which can form the basis of the qubit space. Hence, any state of the
two-dimensional Hilbert space spanned by the basic states
$|L\rangle$ and $|S\rangle$ can be prepared. It has been proved that the time-bin entanglement is a robust form of optical quantum
information, especially for the transmission in optical fibers. In this way, it has been intensively used in quantum communication
over optical fibers and wiveguides, such as the quantum teleportation \cite{time1}, entanglement swapping \cite{swapping1,swapping2} and long-distance entanglement distribution \cite{distribution1,distribution2,distribution3}. In
2002, Thew \emph{et al.} demonstrated the robust time-bin qubits for distributed
quantum communication over $11$ $km$ \cite{distribution1}. Later,
Marcikic \emph{et al.} experimentally realized the distribution of
time-bin entangled qubits over $50$ $km$ of optical fibers \cite{distribution2}. Recently, the group of Inagaki has realized the distribution of time-bin entangled photons over $300$ $km$ of optical fiber \cite{distribution3}. If we send a time-bin qubit $\alpha|L\rangle+\beta|S\rangle$ to different parties, it can creates the single-photon entanglement (SPE). The SPE has important application in the quantum repeater protocols in long-distance quantum communication. For example, in the well known Duan-Lukin-Cirac-Zoller (DLCZ) repeater protocol \cite{SPE1,SPE2}, with one pair
source and one quantum memory at each location, the quantum
repeater can entangle two remote locations A and B. In 2012, Gottesman \emph{et al.} proposed a protocol for
building an interferometric telescope based on the SPE \cite{SPE3}. The protocol has the potential to eliminate the
baseline length limit, and allows the interferometers
with arbitrarily long baselines in principle.

Unfortunately, any kind of entanglement would inevitably suffer the environmental noise during the transmission and storage process, which may cause the entanglement decoherence or photon loss. The entanglement decoherence problem can be combated by the entanglement purification and concentration protocols  \cite{purification1,addpurification2,purification3,purification4,purification8,
purification11,purification12,purification16,purification17,ECP1,ECP2,ECP3,ECP4,ECP5}. On the other hand, due to the photon loss, the entanglement between two distant locations decreases exponentially with the length of the transmission channel. In this way, the photon loss has become the main obstacle for long-distance quantum communication. It will make the pure state mixed with the vacuum state with some probability. In
2009, Ralph and Lund first proposed the concept of the noiseless
linear amplification (NLA) to overcome the exponential
fidelity decay \cite{NLA1}. Since then, many NLA protocols have
been proposed, successively \cite{NLA2,NLA3,NLA4,NLA5,NLA6,NLA7,NLA8,NLA9,NLA10,NLA11,NLA12,NLA13,NLA14,NLA15,nonlinear}. For example, in 2012, Osorio \emph{et al.} experimentally realized the
heralded noiseless amplification based on the single-photon
sources and the linear optics \cite{NLA6}. In the same year, Zhang \emph{et al.} proposed the NLA to protect
the two-mode SPE from photon loss \cite{NLA8}. In 2015, the group of Sheng successfully protect the SPE with imperfect
single-photon source \cite{NLA15}. In the same year, they also proposed the recyclable amplification protocol for the SPE with the weak cross-Kerr nonlinearity. By repeating this amplification, they can effectively increase the fidelity of the output state \cite{nonlinear}. Unfortunately, none of the above amplification protocols can deal with the SPE of the time-bin qubit. Recently, a heralded qubit amplification protocol for the time-bin qubit and polarization qubit is proposed. This protocol is in linear optics, which highlights the simplicity, the stability and potential for fully integrated photonic solutions \cite{timebin}.

Inspired by Ref. \cite{timebin}, in this paper, we will put forward an amplification protocol for the SPE of the time-bin qubit with the help of some auxiliary single photons. In the protocol, we only require some linear optical elements, which makes our protocol can be easily realized under current experimental condition.  The paper is organized as follows, in Sec. 2, we explain the details of the amplification protocol, in Sec. 3, we calculate the amplification factor and approximate success probability and make a discussion and summary.

\section{Heralded amplification for the SPE of the time-bin qubit}
\begin{figure}[!h]%[tpb]
\begin{center}
\includegraphics[width=10cm,angle=0]{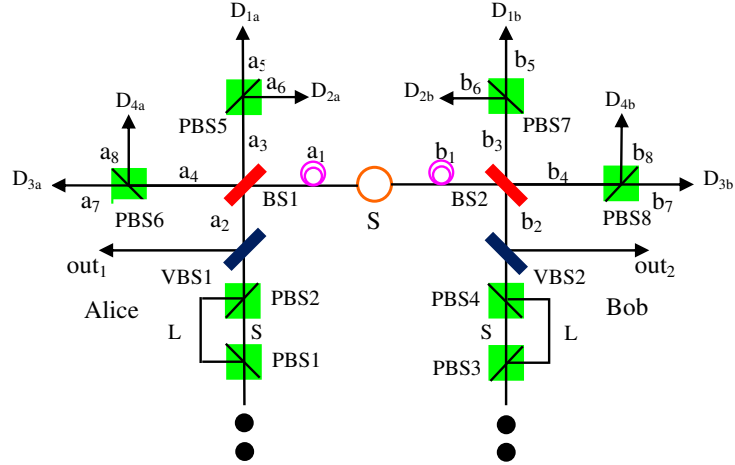}
\caption{The schematic principle of the amplification protocol for the SPE of the time-bin qubit assisted with some auxiliary single photons and the VBSs with the transmission of $t$. }
\end{center}
\end{figure}
In this section, we will explain our amplification protocol for the SPE of the time-bin qubit in detail. We suppose that a time-bin qubit can be written as
\begin{eqnarray}
|\psi\rangle=\alpha|S_{H}\rangle+\beta|L_{V}\rangle
\end{eqnarray}
where $S$ ($L$) stands for the $short$ ($long$) time-bin. The subscript $H$ or $V$ indicates that the the photon is in horizontal polarization or vertical polarization. Here, the polarization is used to label and switch the path of the photon. $\alpha$ and $\beta$ are the entanglement coefficients, where $|\alpha|^{2}+|\beta|^{2}=1$. The time-bin qubit $|\psi\rangle$ is distributed to two parties, say, Alice (A) and Bob (B), which can create a single-photon entangled state as
\begin{eqnarray}
|\Phi\rangle_{AB}=\frac{1}{\sqrt{2}}(|\psi_{a1}0_{b1}\rangle+|0_{a1}\psi_{b1}\rangle).
\end{eqnarray}
Considering the photon loss may occur in the distribution channel, which makes the entanglement mix with the vacuum state, we suppose the single-photon can be completely lost with the probability of $1-\eta$. In this way, the two parties share a mixed state as
 \begin{eqnarray}
\rho_{in}=\eta|\Phi\rangle_{AB}\langle\Phi|+(1-\eta)|vac\rangle\langle vac|.
\end{eqnarray}
In the protocol, we aim to increase the fidelity $\eta$.

The schematic principle of our amplification protocol is shown in Fig. 1. For realizing the amplification, the two parties both need to prepare two auxiliary single-photons, one in $|H\rangle$, and other one in $|V\rangle$. Then, both of them make the auxiliary photons in their hands pass through two polarizing beam splitters (PBSs), say PBS1, PBS2, and PBS3, PBS4, respectively. The PBS can totally transmit the photon in $|H\rangle$ but totally reflect the photon in $|V\rangle$, respectively. Considering the path-length difference, they can create the auxiliary states as $|\varphi\rangle=|S_{H}\rangle\otimes|L_{V}\rangle$.

Next, Alice and Bob make the auxiliary photons in their hands pass through two variable beam splitters (VBSs), which are called VBS1 and VBS2, respectively. Both the two VBSs have the same transmission of $t$. After the VBSs, they can obtain
\begin{eqnarray}
|\varphi_{AB}\rangle&=&(\sqrt{t}|S_{H}0\rangle_{a2out1}+\sqrt{1-t}|0S_{H}\rangle_{a2out1})\otimes(\sqrt{t}|L_{V}0\rangle_{a2out1}+\sqrt{1-t}|0L_{V}\rangle_{a2out1})\nonumber\\
&\otimes&(\sqrt{t}|S_{H}0\rangle_{b2out2}+\sqrt{1-t}|0S_{H}\rangle_{b2out2})\otimes(\sqrt{t}|L_{V}0\rangle_{b2out2}+\sqrt{1-t}|0L_{V}\rangle_{b2out2})\nonumber\\
&=&[t|S_{H}L_{V}\rangle_{a2}+(1-t)|S_{H}L_{V}\rangle_{out1}+\sqrt{t(1-t)}(|S_{H}L_{V}\rangle_{a2out1}+|L_{V}S_{H}\rangle_{a2out1})]\nonumber\\
&\otimes&[t|S_{H}L_{V}\rangle_{b2}+(1-t)|S_{H}L_{V}\rangle_{out2}+\sqrt{t(1-t)}(|S_{H}L_{V}\rangle_{b2out2}+|L_{V}S_{H}\rangle_{b2out2})].
\end{eqnarray}

Combined the initial input state with the auxiliary photon states, the whole state $\rho_{in}\otimes|\varphi_{AB}\rangle$ can be described as follows. It is in the state of $|\Phi\rangle_{AB}\otimes|\varphi_{AB}\rangle$ with the probability of $\eta$, or in the state of $|vac\rangle\otimes|\varphi_{AB}\rangle$ with the probability of $1-\eta$. We first discuss the case of $|\Phi\rangle_{AB}\otimes|\varphi_{AB}\rangle$. The whole photon state can be written as
\begin{eqnarray}
&&|\Phi\rangle_{AB}\otimes|\varphi_{AB}\rangle=\frac{1}{\sqrt{2}}[(\alpha|S_{H}\rangle_{a1}+\beta|L_{V}\rangle_{a1})|0\rangle_{b1}
+|0\rangle_{a1}(\alpha|S_{H}\rangle_{b1}+\beta|L_{V}\rangle_{b1})]\nonumber\\
&\otimes&[t|S_{H}L_{V}\rangle_{a2}+(1-t)|S_{H}L_{V}\rangle_{out1}+\sqrt{t(1-t)}(|S_{H}L_{V}\rangle_{a2out1}+|L_{V}S_{H}\rangle_{a2out1})]\nonumber\\
&\otimes&[t|S_{H}L_{V}\rangle_{b2}+(1-t)|S_{H}L_{V}\rangle_{out2}+\sqrt{t(1-t)}(|S_{H}L_{V}\rangle_{b2out2}+|L_{V}S_{H}\rangle_{b2out2})]\nonumber\\
&=&\frac{1}{\sqrt{2}}[(\alpha|S_{H}\rangle_{a1}+\beta|L_{V}\rangle_{a1})|0\rangle_{b1}
+|0\rangle_{a1}(\alpha|S_{H}\rangle_{b1}+\beta|L_{V}\rangle_{b1})]\nonumber\\
&\otimes&[t|S_{H}L_{V}\rangle_{a2}+(1-t)|S_{H}L_{V}\rangle_{out1}+\sqrt{t(1-t)}(|S_{H}L_{V}\rangle_{a2out1}+|L_{V}S_{H}\rangle_{a2out1})]\nonumber\\
&\otimes&[t|S_{H}L_{V}\rangle_{b2}+(1-t)|S_{H}L_{V}\rangle_{out2}+\sqrt{t(1-t)}(|S_{H}L_{V}\rangle_{b2out2}+|L_{V}S_{H}\rangle_{b2out2})].\label{whole1}
\end{eqnarray}

The two parties make the photons in the $a_{1}a_{2}$ and $b_{1}b_{2}$ modes pass through two $50:50$ beam splitters (BSs), here named BS1 and BS2, respectively. The BSs can make
\begin{eqnarray}
|1\rangle_{a1}&=&\frac{1}{\sqrt{2}}(|1\rangle_{a3}+|1\rangle_{a4}),\quad |1\rangle_{a2}=\frac{1}{\sqrt{2}}(|1\rangle_{a3}-|1\rangle_{a4}),\nonumber\\
|1\rangle_{b1}&=&\frac{1}{\sqrt{2}}(|1\rangle_{b3}+|1\rangle_{b4}),\quad |1\rangle_{b2}=\frac{1}{\sqrt{2}}(|1\rangle_{b3}-|1\rangle_{b4}).
\end{eqnarray}
After that, the state in Eq. (\ref{whole1}) will evolve to
\begin{eqnarray}
|\Phi\rangle_{AB}\otimes|\varphi_{AB}\rangle&\rightarrow&\frac{1}{2}\{(\alpha|S_{H}\rangle_{a3}+\alpha|S_{H}\rangle_{a4}+\beta|L_{V}\rangle_{a3}+\beta|L_{V}\rangle_{a4})\nonumber\\
&\otimes&[\frac{t}{2}(|S_{H}L_{V}\rangle_{a3}-|S_{H}L_{V}\rangle_{a3a4}-|L_{V}S_{H}\rangle_{a3a4}+|S_{H}L_{V}\rangle_{a4})+(1-t)|S_{H}L_{V}\rangle_{out1}\nonumber\\
&+&\frac{\sqrt{t(1-t)}}{\sqrt{2}}(|S_{H}L_{V}\rangle_{a3out1}-|S_{H}L_{V}\rangle_{a4out1}-|L_{V}S_{H}\rangle_{a3out1}-|L_{V}S_{H}\rangle_{a4out1})]\nonumber\\
&\otimes&[\frac{t}{2}(|S_{H}L_{V}\rangle_{b3}-|S_{H}L_{V}\rangle_{b3b4}-|L_{V}S_{H}\rangle_{b3b4}+|S_{H}L_{V}\rangle_{b4})
+(1-t)|S_{H}L_{V}\rangle_{out2}\nonumber\\
&+&\frac{\sqrt{t(1-t)}}{\sqrt{2}}(|S_{H}L_{V}\rangle_{b3out2}-|S_{H}L_{V}\rangle_{b4out2}-|L_{V}S_{H}\rangle_{b3out2}-|L_{V}S_{H}\rangle_{b4out2})]\nonumber\\
&+&(\alpha|S_{H}\rangle_{b3}+\alpha|S_{H}\rangle_{b4}+\beta|L_{V}\rangle_{b3}+\beta|L_{V}\rangle_{b4})\nonumber\\
&\otimes&[\frac{t}{2}(|S_{H}L_{V}\rangle_{a3}-|S_{H}L_{V}\rangle_{a3a4}-|L_{V}S_{H}\rangle_{a3a4}+|S_{H}L_{V}\rangle_{a4})
+(1-t)|S_{H}L_{V}\rangle_{out1}\nonumber\\
&+&\frac{\sqrt{t(1-t)}}{\sqrt{2}}(|S_{H}L_{V}\rangle_{a3out1}-|S_{H}L_{V}\rangle_{a4out1}-|L_{V}S_{H}\rangle_{a3out1}-|L_{V}S_{H}\rangle_{a4out1})]\nonumber\\
&\otimes&[\frac{t}{2}(|S_{H}L_{V}\rangle_{b3}-|S_{H}L_{V}\rangle_{b3b4}-|L_{V}S_{H}\rangle_{b3b4}+|S_{H}L_{V}\rangle_{b4})
+(1-t)|S_{H}L_{V}\rangle_{out2}\nonumber\\
&+&\frac{\sqrt{t(1-t)}}{\sqrt{2}}(|S_{H}L_{V}\rangle_{b3out2}-|S_{H}L_{V}\rangle_{b4out2}-|L_{V}S_{H}\rangle_{b3out2}-|L_{V}S_{H}\rangle_{b4out2})]\}.\label{whole2}
\end{eqnarray}

 Then, the parties make a Bell-state measurement (BSM) for the output photons. In detail, they make the photons in the $a_{3}a_{4}$ and $b_{3}b_{4}$ modes pass through $PBS5$, $PBS6$ and $PBS7$, $PBS8$, respectively, which can make
\begin{eqnarray}
|S_{H}\rangle_{a3(b3)}&\rightarrow&|S_{H}\rangle_{a5(b5)},\quad |L_{V}\rangle_{a3(b3)}\rightarrow|L_{V}\rangle_{a6(b6)},\nonumber\\
|S_{H}\rangle_{a4(b4)}&\rightarrow&|S_{H}\rangle_{a7(b7)},\quad |L_{V}\rangle_{a4(b4)}\rightarrow|L_{V}\rangle_{a8(b8)}.
\end{eqnarray}
After the PBSs, the state in Eq. (\ref{whole2}) will evolve to
\begin{eqnarray}
|\Phi\rangle_{AB}\otimes|\varphi_{AB}\rangle&\rightarrow&\frac{1}{2}\{(\alpha|S_{H}\rangle_{a5}+\alpha|S_{H}\rangle_{a7}+\beta|L_{V}\rangle_{a6}+\beta|L_{V}\rangle_{a8})\nonumber\\
&\otimes&[\frac{t}{2}(|S_{H}\rangle_{a5}|L_{V}\rangle_{a6}-|S_{H}\rangle_{a5}|L_{V}\rangle_{a8}-|L_{V}\rangle_{a6}|S_{H}\rangle_{a7}+|S_{H}\rangle_{a7}|L_{V}\rangle_{a8})+(1-t)|S_{H}L_{V}\rangle_{out1}\nonumber\\
&+&\frac{\sqrt{t(1-t)}}{\sqrt{2}}(|S_{H}\rangle_{a5}|L_{V}\rangle_{out1}-|S_{H}\rangle_{a7}|L_{V}\rangle_{out1}-|L_{V}\rangle_{a6}|S_{H}\rangle_{out1}-|L_{V}\rangle_{a8}|S_{H}\rangle_{out1})]\nonumber\\
&\otimes&[\frac{t}{2}(|S_{H}\rangle_{b5}|L_{V}\rangle_{b6}-|S_{H}\rangle_{b5}|L_{V}\rangle_{b8}-|L_{V}\rangle_{b6}|S_{H}\rangle_{b7}+|S_{H}\rangle_{b7}|L_{V}\rangle_{b8})
+(1-t)|S_{H}L_{V}\rangle_{out2}\nonumber\\
&+&\frac{\sqrt{t(1-t)}}{\sqrt{2}}(|S_{H}\rangle_{b5}|L_{V}\rangle_{out2}-|S_{H}\rangle_{b7}|L_{V}\rangle_{out2}-|L_{V}\rangle_{b6}|S_{H}\rangle_{out2}-|L_{V}\rangle_{b8}|S_{H}\rangle_{out2})]\nonumber\\
&+&(\alpha|S_{H}\rangle_{b5}+\alpha|S_{H}\rangle_{b7}+\beta|L_{V}\rangle_{b6}+\beta|L_{V}\rangle_{b8})\nonumber\\
&\otimes&[\frac{t}{2}(|S_{H}\rangle_{a5}|L_{V}\rangle_{a6}-|S_{H}\rangle_{a5}|L_{V}\rangle_{a8}-|L_{V}\rangle_{a6}|S_{H}\rangle_{a7}+|S_{H}\rangle_{a7}|L_{V}\rangle_{a8})
+(1-t)|S_{H}L_{V}\rangle_{out1}\nonumber\\
&+&\frac{\sqrt{t(1-t)}}{\sqrt{2}}(|S_{H}\rangle_{a5}|L_{V}\rangle_{out1}-|S_{H}\rangle_{a7}|L_{V}\rangle_{out1}-|L_{V}\rangle_{a6}|S_{H}\rangle_{out1}-|L_{V}\rangle_{a8}|S_{H}\rangle_{out1})]\nonumber\\
&\otimes&[\frac{t}{2}(|S_{H}\rangle_{b5}|L_{V}\rangle_{b6}-|S_{H}\rangle_{b5}|L_{V}\rangle_{b8}-|L_{V}\rangle_{b6}|S_{H}\rangle_{b7}+|S_{H}\rangle_{b7}|L_{V}\rangle_{b8})
+(1-t)|S_{H}L_{V}\rangle_{out2}\nonumber\\
&+&\frac{\sqrt{t(1-t)}}{\sqrt{2}}(|S_{H}\rangle_{b5}|L_{V}\rangle_{out2}-|S_{H}\rangle_{b7}|L_{V}\rangle_{out2}-|L_{V}\rangle_{b6}|S_{H}\rangle_{out2}-|L_{V}\rangle_{b8}|S_{H}\rangle_{out2})]\}.\label{whole3}
\end{eqnarray}

Then, the photons in the eight output modes are detected by the single-photon detectors, say, $D_{1a}$, $D_{2a}$, $D_{3a}$, $D_{4a}$, $D_{1b}$, $D_{2b}$, $D_{3b}$, $D_{4b}$, respectively. After the detection, if the detectors $D_{1a}D_{2a}$, $D_{1a}D_{4a}$, $D_{2a}D_{3a}$, or $D_{3a}D_{4a}$ in Alice's location each registers a photon, simultaneously, the detectors $D_{1b}D_{2b}$, $D_{1b}D_{4b}$, $D_{2b}D_{3b}$, or $D_{3b}D_{4b}$ in Bob's location each registers a photon, our protocol will be successful. Otherwise, the protocol is a failure. Therefore, there are sixteen successful detection results in total, which are displayed in Table 1 as follows.
\begin{center}
\tabcolsep=8pt  %%% 此参数用于调整表的整体宽度
\small
\renewcommand\arraystretch{1}  %% 调整表内行与行之间的纵向距离
\begin{minipage}{12cm}
\small{Table 1. The successful detection results of our protocol, where $\circ$ means our protocol is successful under this detection result.}
\end{minipage}
\vglue5pt
\begin{tabular}{| c | c | c | c | c |}
\hline
  &$D_{1a}D_{2a}$ & $D_{1a}D_{4a}$ &$D_{2a}D_{3a}$& $D_{3a}D_{4a}$ \\
\hline
  $D_{1b}D_{2b}$& $\circ$ & $\circ$ & $\circ$ & $\circ$ \\
  $D_{1b}D_{4b}$& $\circ$ & $\circ$ & $\circ$ & $\circ$ \\
  $D_{2b}D_{3b}$& $\circ$ & $\circ$ & $\circ$ & $\circ$ \\
  $D_{3b}D_{4b}$& $\circ$ & $\circ$ & $\circ$ & $\circ$ \\
\hline
\end{tabular}
\end{center}
\vspace*{2mm}

We take the detection result of $D_{1a}D_{2a}D_{1b}D_{2b}$ for example. In Eq. (\ref{whole2}), the four items $\frac{-\alpha t\sqrt{t(1-t)}}{4\sqrt{2}}|S_{H}\rangle_{a5}|L_{V}\rangle_{a6}|S_{H}\rangle_{b5}|L_{V}\rangle_{b6}|S_{H}\rangle_{out1}$, $\frac{\beta t\sqrt{t(1-t)}}{4\sqrt{2}}|L_{V}\rangle_{a6}|S_{H}\rangle_{a5}|S_{H}\rangle_{b5}|L_{V}\rangle_{b6}|L_{V}\rangle_{out1}$, $\frac{-\alpha t\sqrt{t(1-t)}}{4\sqrt{2}}|S_{H}\rangle_{b5}|L_{V}\rangle_{b6}|S_{H}\rangle_{a5}|L_{V}\rangle_{a6}|S_{H}\rangle_{out2}$, and $\frac{\beta t\sqrt{t(1-t)}}{4\sqrt{2}}|L_{V}\rangle_{b6}|S_{H}\rangle_{b5}|S_{H}\rangle_{a5}|L_{V}\rangle_{a6}|L_{V}\rangle_{out2}$ will lead to the detectors $D_{1a}D_{2a}D_{1b}D_{2b}$ each registers one photon. In this way, if the parties obtain this detection result, the state in Eq. (\ref{whole3}) will collapse to
\begin{eqnarray}
|\Phi'_{1}\rangle_{AB}=(-\alpha|S_{H}\rangle_{out1}+\beta|L_{V}\rangle_{out1})|0\rangle_{out2}
+|0\rangle_{out1}(-\alpha|S_{H}\rangle_{out2}+\beta|L_{V}\rangle_{out2}),\label{result1}
\end{eqnarray}
with the probability of $\frac{t^{3}(1-t)}{16}$.

Finally, the parties only need to perform a phase-flip operation on the photon in the output1 or output2 mode, they can change $|\Phi'_{1}\rangle_{AB}$ in Eq. (\ref{result1}) to
\begin{eqnarray}
|\Phi_{1}\rangle_{AB}=(\alpha|S_{H}\rangle_{out1}+\beta|L_{V}\rangle_{out1})|0\rangle_{out2}
+|0\rangle_{out1}(\alpha|S_{H}\rangle_{out2}+\beta|L_{V}\rangle_{out2}),\label{result2}
\end{eqnarray}
which has the same form of $|\Phi\rangle_{AB}$.

If the parties obtain one of the other fifteen detection results in Table 1, they will finally obtain the same result of $|\Phi_{1}\rangle_{AB}$ in Eq. (\ref{result2}). Therefore, the total success probability to obtain $|\Phi_{1}\rangle_{AB}$ can be written as $P_{1}=16\times\frac{t^{3}(1-t)}{16}=t^{3}(1-t)$.

On the other hand, if the initial input state is the vacuum state, after the auxiliary photons pass through BS1 and BS2, respectively, the parties can obtain
\begin{eqnarray}
|vac\rangle\otimes|\varphi_{AB}\rangle&\rightarrow&[\frac{t}{2}(|S_{H}L_{V}\rangle_{a3}-|S_{H}L_{V}\rangle_{a3a4}-|L_{V}S_{H}\rangle_{a3a4}+|S_{H}L_{V}\rangle_{a4})+(1-t)|S_{H}L_{V}\rangle_{out1}\nonumber\\
&+&\frac{\sqrt{t(1-t)}}{\sqrt{2}}(|S_{H}L_{V}\rangle_{a3out1}-|S_{H}L_{V}\rangle_{a4out1}-|L_{V}S_{H}\rangle_{a3out1}-|L_{V}S_{H}\rangle_{a4out1})]\nonumber\\
&\otimes&[\frac{t}{2}(|S_{H}L_{V}\rangle_{b3}-|S_{H}L_{V}\rangle_{b3b4}-|L_{V}S_{H}\rangle_{b3b4}+|S_{H}L_{V}\rangle_{b4})
+(1-t)|S_{H}L_{V}\rangle_{out2}\nonumber\\
&+&\frac{\sqrt{t(1-t)}}{\sqrt{2}}(|S_{H}L_{V}\rangle_{b3out2}-|S_{H}L_{V}\rangle_{b4out2}-|L_{V}S_{H}\rangle_{b3out2}-|L_{V}S_{H}\rangle_{b4out2})].
\label{whole4}
\end{eqnarray}

Then the parties make the output photons pass through the PBSs. The state in Eq. (\ref{whole4}) will evolve to
\begin{eqnarray}
|vac\rangle\otimes|\varphi_{AB}\rangle&\rightarrow&[\frac{t}{2}(|S_{H}\rangle_{a5}|L_{V}\rangle_{a6}-|S_{H}\rangle_{a5}|L_{V}\rangle_{a8}-|L_{V}\rangle_{a6}|S_{H}\rangle_{a7}+|S_{H}\rangle_{a7}|L_{V}\rangle_{a8})
+(1-t)|S_{H}L_{V}\rangle_{out1}\nonumber\\
&+&\frac{\sqrt{t(1-t)}}{\sqrt{2}}(|S_{H}\rangle_{a5}|L_{V}\rangle_{out1}-|S_{H}\rangle_{a7}|L_{V}\rangle_{out1}-|L_{V}\rangle_{a6}|S_{H}\rangle_{out1}-|L_{V}\rangle_{a8}|S_{H}\rangle_{out1})]\nonumber\\
&\otimes&[\frac{t}{2}(|S_{H}\rangle_{b5}|L_{V}\rangle_{b6}-|S_{H}\rangle_{b5}|L_{V}\rangle_{b8}-|L_{V}\rangle_{b6}|S_{H}\rangle_{b7}+|S_{H}\rangle_{b7}|L_{V}\rangle_{b8})
+(1-t)|S_{H}L_{V}\rangle_{out2}\nonumber\\
&+&\frac{\sqrt{t(1-t)}}{\sqrt{2}}(|S_{H}\rangle_{b5}|L_{V}\rangle_{out2}-|S_{H}\rangle_{b7}|L_{V}\rangle_{out2}-|L_{V}\rangle_{b6}|S_{H}\rangle_{out2}-|L_{V}\rangle_{b8}|S_{H}\rangle_{out2})].
\label{whole5}
\end{eqnarray}

We will prove that under the sixteen successful cases, the parties will finally obtain the $|vac\rangle$ state. We also take the detection result of $D_{1a}D_{2a}D_{1b}D_{2b}$ for example. It can be found that only the item $\frac{t^{2}}{4}|S_{H}\rangle_{a3}|L_{V}\rangle_{a3}|S_{H}\rangle_{b3}|L_{V}\rangle_{b3}$ will lead to the detectors $D_{1a}D_{2a}D_{1b}D_{2b}$ each registers one photon. If the detection result is $D_{1a}D_{2a}D_{1b}D_{2b}$, the state in Eq. (\ref{whole5}) will finally evolve to the vacuum state with the probability of $\frac{t^{4}}{16}$. Certainly, if the parties obtain any one of the other fifteen detection results in Table 1, they can also obtain the same result. Therefore, the total success probability to obtain the vacuum state is $P_{2}=16\times\frac{t^{4}}{16}=t^{4}$.

According to the description above, the total success probability ($P_{t}$) of our protocol can be written as
\begin{eqnarray}
P_{t}=\eta P_{1}+(1-\eta)P_{2}=\eta t^{3}(1-t)+ (1-\eta)t^{4}=(1-2\eta)t^{4}+\eta t^{3}.
\end{eqnarray}
When the protocol is successful, the parties can obtain a mixed state as
\begin{eqnarray}
\rho_{out}=\eta'|\Phi_{1}\rangle_{AB}\langle\Phi_{1}|+(1-\eta')|vac\rangle\langle vac|
\end{eqnarray}
with the fidelity
\begin{eqnarray}
\eta'=\frac{\eta t^{3}(1-t)}{\eta t^{3}(1-t)+(1-\eta)t^{4}}=\frac{\eta (1-t)}{\eta (1-t)+(1-\eta)t}.\label{fidelity}
\end{eqnarray}

It can be found that the fidelity $\eta'$ of the new mixed state has nothing to do with the entanglement coefficients $\alpha$ and $\beta$, but it only depends on the fidelity $\eta$ of the initial input mixed state and the transmission $t$ of the VBSs. We design the amplification factor as \begin{eqnarray}
g\equiv\frac{\eta'}{\eta}=\frac{1-t}{\eta (1-t)+(1-\eta)t}. \label{g}
\end{eqnarray}
For realizing the amplification, we require $\eta'>\eta$, that is, $g>1$. It can be calculated that $g>1$ under the case of $t<\frac{1}{2}$. In this way, by providing suitable VBSs with $t<\frac{1}{2}$, we can complete the amplification task.

\section{Discussion and conclusion}
\begin{figure}[!h]%[tpb]
\begin{center}
\includegraphics[width=7cm,angle=0]{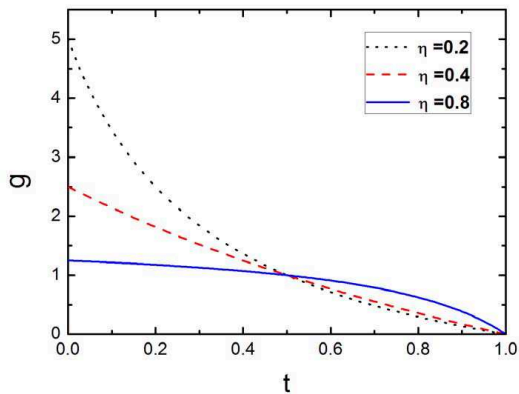}
\caption{The amplification factor $g$ as a function of the transmission $t$ of the VBSs under the initial fidelity $\eta=0.2$, 0.4, and 0.8, respectively. All three curves pass through the same point with $t=0.5$.}
\end{center}
\end{figure}

\begin{figure}[!h]%[tpb]
\begin{center}
\includegraphics[width=7cm,angle=0]{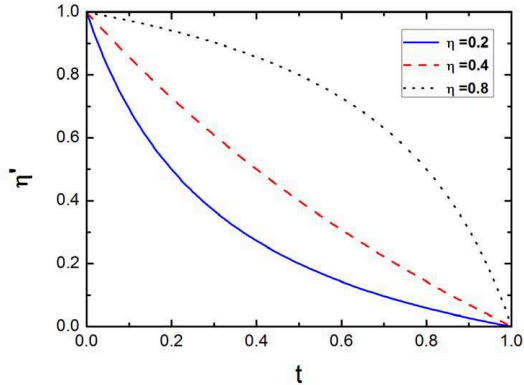}
\caption{The fidelity $\eta'$ of the distilled new mixed state altered with the transmission $t$ of the VBSs under the initial fidelity $\eta=0.2$, 0.4, and 0.8, respectively. }
\end{center}
\end{figure}

In the paper, we propose a simple and efficient amplification protocol for protecting the single-photon entangled state of the time-bin qubit. In the protocol, one time-bin qubit is shared by two parties, which creates a single-photon entangled state. Due to the photon loss, the entangled state would be mixed with the vacuum state. For realizing the amplification, each of the two parties requires to prepare two single photons with the polarization of $|H\rangle$ and $|V\rangle$. With the help of the PBSs, the polarization modes can accompany two temporal modes, respectively. Then, each party makes the auxiliary photons in his or her hand pass through a VBS with the transmission of $t$. Subsequently, each party makes the photons enter the $BS$, and make the BSM for the photons in the eight output modes. According to the BSM result, the parties can distill a new mixed state with the similar form of the initial mixed state. Under the case that $t<\frac{1}{2}$, we can obtain the amplification factor $g\equiv\frac{\eta'}{\eta}>1$ and realize the amplification. Our protocol has three obvious advantages.
First, we only require one pair of the single-photon entangled state. As the entanglement source
is quite precious, our protocol is quite economical. Second, the encoded time-bin feature can be perfectly contained. Third, we only require the linear optical
elements, which makes our protocol can be realized in current experimental
conditions.

The the amplification factor $g$ and the fidelity $\eta'$ of the distilled new mixed state as a function of the transmission $t$ of the VBSs are shown in Fig. 2 and Fig. 3, respectively. In Fig. 2, and Fig. 3, both the values of $g$ and $\eta'$ reduce with the growth of $t$. All curves in Fig. 2 pass through the same point with $t=\frac{1}{2}$. Under $t=\frac{1}{2}$, we can obtain $g=1$ and $\eta'=\eta$ under any initial coefficient $\eta$. Actually, under $t=\frac{1}{2}$, the VBS becomes the BS and our protocol is analogous to an entanglement swapping protocol. Under $t<\frac{1}{2}$, we can obtain $g>1$ and $\eta'>\eta$. Combined with Eq. (\ref{fidelity}) and Eq. (\ref{g}), we can obtain under $t\rightarrow 0$, $\eta'\rightarrow 1$ and $g\rightarrow\frac{1}{\eta}$. In this way, for obtaining high fidelity, the parties require to choose the VBS with small transmission.

\begin{figure}[!h]%[tpb]
\begin{center}
\includegraphics[width=7cm,angle=0]{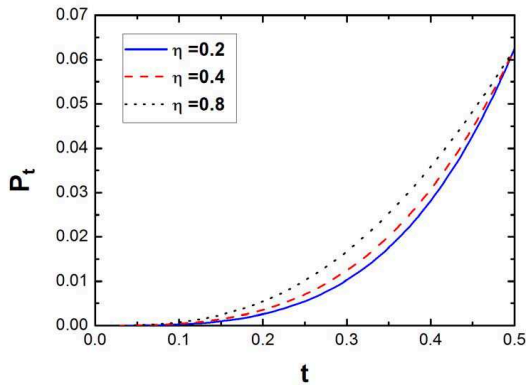}
\caption{The total success probability $P_{t}$ of our amplification protocol as a function of the transmission $t$ of the VBSs under the initial fidelity $\eta=0.2$, 0.4, and 0.8, respectively. }
\end{center}
\end{figure}

On the other hand, we calculate the total success probability $P_{t}$ of our protocol as a function of the transmission $t$ of the VBSs under three different initial $\eta=0.2$, $0.4$, and $0.8$, respectively. As shown in Fig. 4, the value of $P_{t}$ mainly depends on $t$ and the value of $\eta$ affects $P_{t}$ slightly. In all the three curves, $P_{t}$ increases with the growth of $t$. Under $t\in [0,\frac{1}{2}]$, $P_{t}$ will get the maximum value of $\frac{1}{16}$ when $t=\frac{1}{2}$. When $t\rightarrow 0$, $P_{t}\rightarrow 0$. In this way, in the practical application, we need to consider both the fidelity and success probability factors, simultaneously, and choose the VBSs with suitable transmission.

Finally, we discuss the experimental realization of our protocol. The VBS is the key element of the protocol. The VBS is a common linear optical
element in current technology. In our protocol, for realizing the amplification, we require to use the VBS with $t<\frac{1}{2}$. In 2012, the group of Osorio reported their experimental results
about the heralded photon amplification for quantum communication
with the help of the VBS \cite{NLA6}. In their amplification experiment, they successfully adjusted the splitting
ratio of VBS from $50:50$ to $90:10$ to increase the visibility from
$46.7\pm 3.1\%$ to $96.3\pm 3.8\%$. Based on their experimental result, our requirement for the VBS can be easily realized. On the other hand, we also require the sophisticated single photon detectors to exactly distinguish
the single photon in each output modes. The single photon detection has been a challenge under current experimental
conditions, for the quantum decoherence effect of the photon detector \cite{photonefficiency}. Lita \emph{et al.} reported their experimental
result about the near-infrared single-photon detection. They showed the photon detection efficiency $\eta_{p}$ at 1556 $nm$ can reach $95\% \pm 2\%$ \cite{photonefficiency1}.

In conclusion, we demonstrate a simple and effective amplification protocol for protecting the single-photon entangled state of the time-bin qubit. In the protocol, we only require one pair of the single-photon entangled state, which makes our protocol economical. With the help of some auxiliary single photons and the linear optical elements, such as VBSs, BSs, and PBSs, the fidelity of the single-photon entangled state can be increased when the transmission $t$ of the VBSs satisfy $t<\frac{1}{2}$. Moreover, the encoded time-bin feature can be well contained. This protocol can be realized under current experimental condition, and it may be useful in current and future quantum information processing.

\section*{ACKNOWLEDGEMENTS} This work was supported by the National Natural Science Foundation
of China under Grant  Nos. 11474168 and 61401222, the Natural Science Foundation of Jiangsu province under Grant No. BK20151502
, the Qing Lan Project in Jiangsu Province, and A Project
Funded by the Priority Academic Program Development of Jiangsu
Higher Education Institutions.

\end{document}